\documentclass[twocolumn,aps,pra,amsmath,amssymb,floatfix]{revtex4}
\usepackage{dcolumn}
\usepackage{graphicx,xcolor}
\usepackage{hyperref}
\usepackage{soul}

\newcommand{\REV}[1]{#1} 

\begin{document}

\title{Femtosecond laser micromachining for integrated quantum photonics}

\author{Giacomo Corrielli$^{1,2}$}
\author{Andrea Crespi$^{2,1}$}
\author{Roberto Osellame$^{1,2,}$}
\email{roberto.osellame@cnr.it}

\affiliation{$^1$Consiglio Nazionale delle Ricerche, Istituto di Fotonica e Nanotecnologie, piazza Leonardo da Vinci 32, 20133 Milano, Italy}
\affiliation{$^2$Dipartimento di Fisica, Politecnico di Milano, piazza Leonardo da Vinci 32, 20133 Milano, Italy}

\date{\today}

\begin{abstract}
Integrated quantum photonics, i.e. the generation, manipulation and detection of quantum states of light in integrated photonic chips, is revolutionizing the field of quantum information in all applications, from communications to computing. Although many different platforms are being currently developed, from silicon photonics to lithium niobate photonic circuits, none of them has shown the versatility of femtosecond laser micromachining (FLM) in producing all the components of a complete quantum system, encompassing quantum sources, reconfigurable state manipulation, quantum memories and detection. It is in fact evident that FLM has been a key enabling tool in the first-time demonstration of many quantum devices and functionalities.
Although FLM cannot achieve the same level of miniaturization of other platforms, it still has many unique advantages for integrated quantum photonics. In particular, in the last five years, FLM has greatly expanded its range of quantum applications with  several scientific breakthroughs achieved. For these reasons, we believe that a review article on this topic is very timely and could further promote the development of this field by convincing end-users of the great potentials of this technological platform and by stimulating more research groups in FLM to direct their efforts to the exciting field of quantum technologies.
\end{abstract}

\maketitle
\section{Introduction}

Great expectations are currently placed on the second quantum revolution \cite{Dowling03}, where quantum technologies promise to dramatically change the way we process information for practical applications. This revolution encompasses all aspects of information, from how we retrieve it with sensing, to how we elaborate it through computing and simulation, to the way we transfer it with communications. 

The focus of the second quantum revolution is mainly on developing new quantum-enabled technologies, in contrast to the first one that mainly aimed at developing new science (to the extent that the two aspects can be considered separately). The emphasis put on practical applications of quantum devices is important to understand the global effort specifically devoted to advancing the technological aspects. In fact, system scalability and engineering are the most important keywords in evaluating and selecting possible quantum platform. The range of physical systems that are currently available to implement the qubit (the elementary unit of quantum information) are several and very different. The most advanced ones include superconducting circuits, trapped ions or atoms, defects in various materials, and photons. Each of them has specific advantages and disadvantages and lends itself particularly to a specific task in quantum information. 

Photonics has unique properties for quantum communications \cite{Gisin07}, but also has great potentials in sensing \cite{pira18}, simulation \cite{aspuru12} and computing \cite{rudolph17}. Although amazing results have been achieved with extremely complex bulk optics set-up \cite{Zhong20}, user-friendly, scalable, portable and reliable photonic devices can only be achieved with integrated optics. The latter is a well developed approach for applications in classical information and could thus be leveraged for a rapid growth of quantum technologies. 

However, different emphasis is put on some aspects of the photonic circuits by classical and quantum operation. In particular, photon losses (that can be easily compensated classically by optical amplifiers) are a key aspect for quantum photonics, where the no-cloning theorem means that some information is irreversibly destroyed with each lost photon. In addition, specific components as single photon sources, memories and detectors need to be developed for a complete quantum photonic system. These aspects motivate on-going and future research in integrated quantum photonics. 

The aim of this review is not to cover the whole field of integrated quantum photonics, which has been the topic of a recent review paper \cite{wang20}, but to summarize the contribution given to this important field by a specific microfabrication technology, i.e. femtosecond laser micromachining (FLM) of transparent materials \cite{osellame12}. In fact, this technological platform has produced a wealth of new and significant contributions to the field since the first results discussed in a previous review \cite{meany15}.

The article is organized in the following way: Section 2 gives a broader view on integrated quantum photonics, describing the main architecture of a quantum photonic system and the most relevant competitors to FLM for its implementation. Section 3 focuses on FLM and its processing possibilities, from waveguide writing, to selective removal of material, to additive manufacturing at the nano/micro-scale with two-photon polymerization. This unique portfolio of processing capabilities (that can be easily combined) is implemented with the same experimental setup and can enable unique quantum devices that are detailed in the following sections. Section 4 focuses on the use of FLM to directly fabricate single photon sources in different materials. If a source is produced with a different technology, FLM can still be useful in devising an efficient interconnection of the source with integrated photonic circuits as described in Section 5. Section 6 deals with the most exploited application of FLM in quantum photonics, i.e. the fabrication of complex integrated photonic circuits for manipulation of quantum states of light, both in static devices and in dynamically-reconfigurable ones. Integrated quantum memories, discussed in Section 7, are very important components that received a significant boost by the adoption of FLM for rare-earth-doped crystals. Finally, in Section 8 we discuss the perspectives of FLM in combining single-photon detectors with integrated photonic circuits.

\section{Integrated quantum photonics}

Photons are very promising carriers of quantum information, in fact they can operate at room temperature and can propagate for long distances with very limited interactions with the environment, thus preserving the quantum state for a long time. On the other hand, their main limitations are that they can be lost along their propagation and that they are mainly suitable for linear manipulation (until strong interactions between individual photons become easy to implement \cite{chang14}).

Integrated quantum photonics is the synergic combination of quantum photonics with integrated optics \cite{Politi08,wang20}. It enables dramatic improvements with respect to bulk optic quantum experiments in terms of compactness, scalability, control and phase stability. In addition, integrated quantum photonics enables a seamless connection with optical fibers yielding robust and alignment-free devices that can be field-deployable. 

 The importance of integrated quantum photonics is evident for quantum communications where their suitability as 'flying qubits' is unique, but they are also finding increasing importance in all other quantum information tasks. Recent work has shown the possibility to simulate vibrational quantum dynamics of molecules in integrated photonic circuits \cite{sparrow18}. Quantum computation hardware has also been demonstrated in integrated photonics, both at the level of probabilistic two-qubit quantum gates \cite{Politi08,Crespi11,zhang19,zhang21}, as well as complex reconfigurable linear photonic processors \cite{Carolan15,Taball19}. In particular, different variants of measurement-based quantum computation hold promise for fault-tolerant photonic quantum computers \cite{takeda19,bourassa21}.
 
 It is worth noting that even when photons are not used as qubits, they are still relevant actors in the quantum architecture, for example in the read out of the state of the quantum system, being it an ion, an atom or a superconducting circuit; also in this case integrated photonics can play an important role \cite{mehta16,leung19}.

\subsection{Architecture for a universal integrated quantum photonic system}

\begin{figure}
    \centering
    \includegraphics{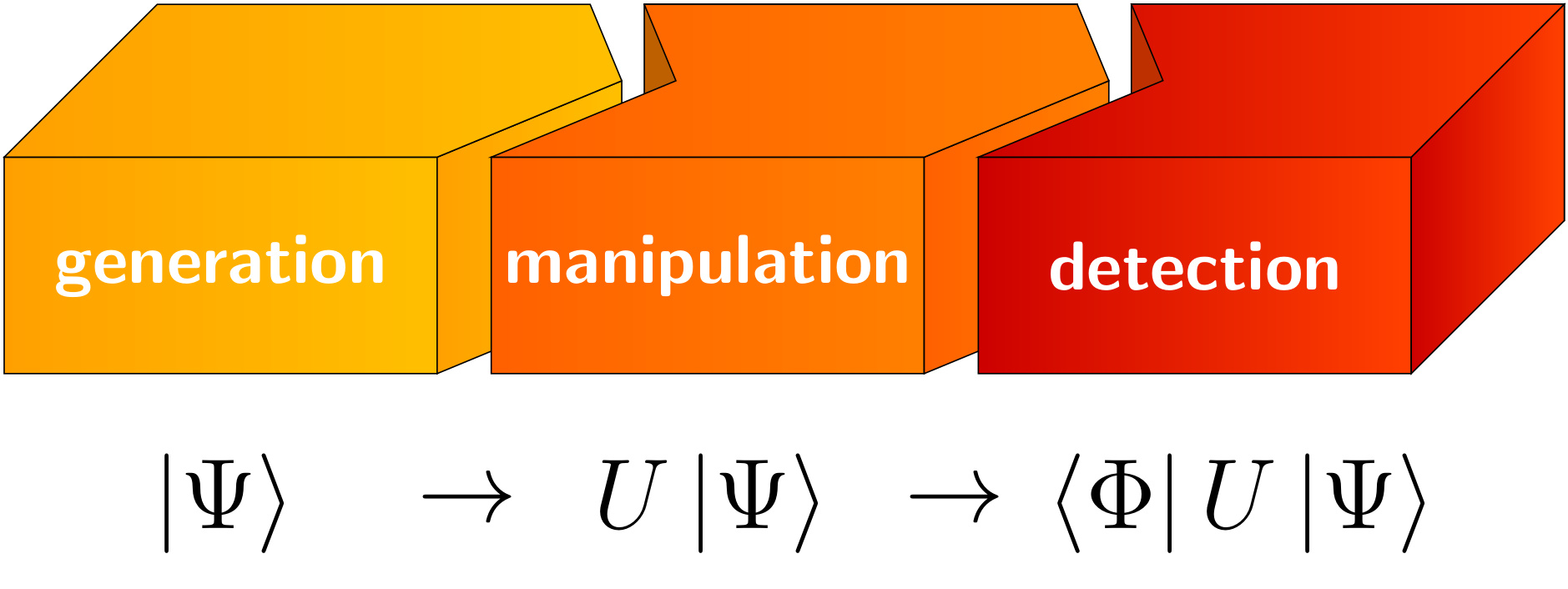}
    \caption{A quantum photonic system is typically built of three blocks. A first block generates a quantum state $\left|\Psi\right\rangle$ of one or more photons, which is useful for further elaboration. A second block manipulates the state, most often by means of a linear unitary transformation $U$. The final block performs the detection, which equals to measuring the projection on the elaborated state on one or more quantum states $\left|\Phi\right\rangle$.}
    \label{fig:blocks}
\end{figure}

For most, if not all, applications, a quantum photonic system can be schematized into three main blocks (Fig.~\ref{fig:blocks}): the block for the generation of single photons or more complex quantum states of light, the block for the manipulation of the information encoded in such quantum states (where the actual quantum protocol is implemented) and a final block for \REV{the} 
measurement through photodetectors. All three block are undergoing a process of miniaturization and integration to improve their portability and scalability. In particular, a significant effort has been devoted to the integration of the second block, where the information is manipulated according to the quantum protocol. 

A current limitation of quantum photonics is that only linear manipulation of the quantum state is typically viable, although nonlinearities can be simulated with measurements of ancillary photons in linear circuits \cite{knill01}. Nevertheless, many tasks can be implemented with linear operations, as well as reaching a computational hardness \cite{Aaron11} that recently led to a quantum advantage demonstration \cite{Zhong20}. In particular, the time evolution of a quantum system under a given Hamiltonian gives rise to a unitary transformation of the initial state that can be accurately mapped in a linear photonic circuit. 
Thanks to the fine control of the photonic circuit properties, quantum devices implementing specific unitary transformations can be fabricated, with the possibility of having different devices for each quantum manipulation task. However, the main trend nowadays is to produce fully reconfigurable photonic circuits to implement multiple, arbitrary unitary transformations in the same circuit.

A final comment on the overall architecture, points to the two current alternatives to combine the three blocks in a complete quantum system. One possibility is to combine all of them monolithically on a single substrate with significant advantages in the long-term for scalability and mass production, but with greater technical difficulties in optimizing all the functionalities on the same material. A second possibility is the modular approach, where each block is developed and optimized in a dedicated module and then the different modules are optically interconnected. This may prove as an advantageous approach in the short term as it can harvest immediately all the results achieved on the different functionalities without having to worry about how to transfer them on the same technological platform. On the other hand, assembly of these modular systems may be complicated and not so easily scalable to very complex devices.   

\subsection{Available integrated photonic platforms}

Miniaturization of quantum photonic devices in integrated optics can leverage from all the work done in classical photonics for other applications. Different integrated photonic platforms have been developed and each of them is being explored for its potential in quantum information applications. Each platform has advantages and disadvantages and thus no clear winner is yet established in integrated quantum photonics. Silicon photonics \cite{silverstone14}, silicon nitride \cite{Taball19}, and silica on silicon \cite{Politi08} are well-consolidated platforms, listed according to a decreasing level of miniaturization. Nonclassical single photon sources can be embedded on-chip exploiting nonlinear four-wave-mixing processes \cite{silverstone14,harris14} and superconducting nanowire detectors have been integrated on a few of these platforms \cite{mattioli20}. Reconfigurability of the photonic circuits is mainly achieved with thermal phase shifters, providing stable but slow phase control. Lithium niobate is another interesting platform as it provides second order nonlinearity and electro-optic properties. Photon sources based on parametric down conversion have been demonstrated on this material \cite{krapick13}, as well as very fast phase control \cite{bonneau12}. This last platform is also gaining significant interest in a new form, i.e. lithium niobate on insulator (LNOI), where the above properties are combined with an increased miniaturization capability \cite{boes18}. \REV{A further integrated photonic platform is femtosecond laser micromachining (FLM) in transparent materials \cite{gattass08,osellame12,meany15}. Although less extreme in terms of device miniaturization if compared to lithographic processes, FLM is a powerful microfabrication technology which shows the distinctive features of three-dimensional capabilities, versatility in terms of kind of processable materials, and rapid prototyping. In fact, low costs and fast fabrication turnaround times of FLM often allow designing the devices by an iterative process of fabrication and characterization, rather than relying on complex numerical simulation. This platform has enabled many breakthroughs in quantum photonics and it will be discussed more in details in the following sections.}

\section{Femtosecond laser micromachining}

In FLM, femtosecond laser pulses are exploited to machine transparent dielectric materials with high spatial resolution in the three dimensions. To the purpose, the femtosecond laser beam is focused by suitable optics (e.g. an aspheric lens or a microscope objective) on the surface or into the bulk of the transparent material \REV{(Fig.~\ref{fig:figureFLM}a)}. The high peak intensity, enabled by the combination of short pulse length and tight focusing, triggers nonlinear absorption processes in the transparent material, which can result in a permanent and localized modification around the focal region. 
\REV{Three-dimensional machining is achieved by translating the substrate with respect to the laser focus during irradiation, typically by means of high-precision translation stages. 

The microscopic modification processes are complex, and originate from an intricate interplay among different phenomena, which include electronic excitation or relaxation, and thermal effects \cite{gattass08}.
Indeed, modifications can be of different nature, encompassing refractive index variations, chemical/structural changes, ablation and creation of micro-voids, depending on the specific material and of on the adopted irradiation parameters. Generally, for increasing pulse energy, one observe the passage from a regime of smooth modification of the optical properties, to the creation of more invasive structural changes, ending with cracks and voids if the deposited energy is too high.

It is also noted that, while the nonlinear absorption phenomena that trigger the modification process occurs well confined in the focal region, the size of the modified region can be quite larger, because of the influence of thermal diffusion or accumulation \cite{eaton08}.
}

We will \REV{now} discuss in more details the processing portfolio of FLM, from waveguide writing to 3D structuring, to two-photon polymerization, and will then conclude emphasising how the specific features of FLM can be beneficial for integrated quantum photonics. 

\subsection{Waveguide writing}

The possibility of producing positive refractive index changes in glass by means of FLM for writing optical channel waveguides has been originally discovered by Davis et al. in 1996 \cite{davis96}. Since then, numerous research groups worldwide have extensively worked for improving the guiding features of laser-written waveguides in many transparent materials, ranging from various kind of glasses, to dielectric crystalline substrates \cite{osellame12, chen14}.

In the femtosecond-laser direct-writing process, waveguides are defined by translating the substrate at constant speed with respect to the laser focus, according to arbitrary three-dimensional paths. 
\REV{The optical performance of the fabricated waveguides depends in a non-trivial way on the irradiation parameters and on the specific substrate. In general, a higher deposited energy (e.g., via a lower translation speed, a higher pulse energy or a higher repetition rate) is typically associated with a larger waveguide cross-section and a higher refractive-index difference of the modified region with respect to the pristine material.}
Typical waveguides fabricated by FLM have a transverse cross section of few $\mu$m$^2$ and show a refractive index change between core and cladding in the order of $10^{-2}-10^{-3}$, which reflects in their excellent connectivity with standard optical fibers. 

\begin{figure*}
    \centering
    \includegraphics{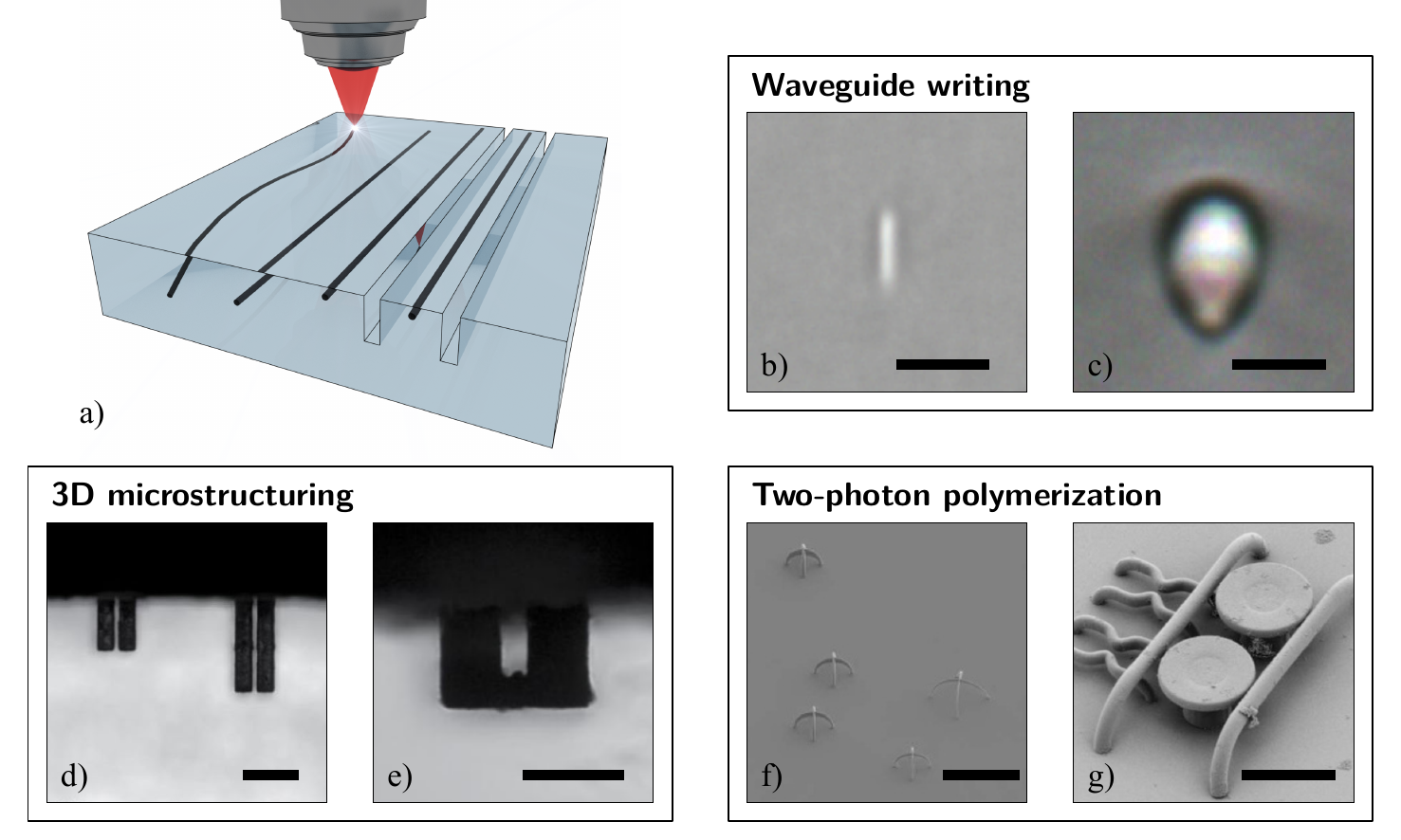}
    \caption{Graphical summary of the capabilities of FLM. Irradiation of a transparent substrate with a focused femtosecond laser beam, to produce buried modified tracks, is represented pictorially in (a); boxes report examples of the different available fabrication processes (b-g). For direct waveguide writing, microscope pictures of FLM waveguides in two different materials are reported: (b) Single-mode waveguide operating in the visible, realized in fused silica substrate by a single femtosecond laser scan at 50~kHz pulse repetition rate (authors' own work, scalebar is 10~$\mu$m); (c) Single-mode waveguide operating at 1550~nm, inscribed in Eagle2000 (Corning Inc.) alumino-borosilicate glass by multiple overlapped laser scans at 5~MHz, followed by thermal annealing (adapted from Ref.~\cite{arriola13}, scalebar is 10~$\mu$m). For 3D microstructuring, cross-sections of different excavated structures, produced by water-assisted laser ablation in EagleXG (Corning Inc.) alumino-borosilicate glass, are reported (both adapted from Ref.~\cite{Cecca20}): (d) Microtrenches realized on both sides of an optical waveguide (scalebar is 200~$\mu$m); (e) U-shaped hollow microstructure, surrounding a thin glass beam where an optical waveguide has also been inscribed (scalebar is 100~$\mu$m). For two-photon polymerization, pictures taken with an electronic microscope of fabricated microstructures are shown: (f) Crossed-arc polymeric waveguides, each crossing contains one pre-characterized single-photon emitter (adapted from Ref.~\cite{shi16}, scalebar is 20~$\mu$m); (g) 3D structure containing several key photonic elements, such as waveguides, couplers
and microdisc resonators (adapted from Ref.~\cite{schell2013}, scalebar is 10~$\mu$m).
    }
    \label{fig:figureFLM}
\end{figure*}

Waveguide inscription in glass is performed by writing directly the waveguide core, i.e. producing a positive refractive index change with the laser irradiation (type I waveguide, see Fig.~\ref{fig:figureFLM}b,c). The most employed substrates, especially for integrated quantum photonics applications, are pure fused silica and commercial borosilicate glasses, e.g. Corning EagleXG and Eagle2000 or Schott AF32. Single mode waveguides have been fabricated in these materials covering the whole visible and near infrared spectrum of light, from 400 nm, up to the telecom C-band, demonstrating propagation losses in the range 0.1-1 dB/cm and negligible bending losses for radii above 30 mm \cite{will02, day21, ferna20}. The most favorable inscription regime for the processing of pure fused silica is with low repetition rates, from 1 kHz to 100 kHz, and low scan speeds (up to few mm/s). Single-scan waveguides fabricated in fused silica typically present a strongly elongated cross section and a moderate degree of modal birefringence, up to 10$^{-4}$. Multiple and partially overlapped scans can be employed for engineering the waveguide cross section, thus producing more circular intensity profiles of the guided mode \cite{nasu05}. For the fabrication of waveguides in borosilicate glasses, instead, higher repetition rates (500 kHz to 5 MHz), and scan speeds from 10 mm/s to 100 mm/s are preferred. In this regime, thermal accumulation effects importantly contribute to the material modification \cite{eaton05,eaton08}, resulting in a more symmetric waveguide cross section and low birefringence (in the order of 10$^{-5}$), mostly given by the residual mechanical stress accumulated in the substrate during the fabrication. Recently, it has been demonstrated that a thermal annealing treatment performed on borosilicate waveguides after fabrication improves the optical mode confinement, allowing for radii of curvature of 10 mm \cite{arriola13}, and quench waveguide birefringence almost completely (in the order of 10$^{-6}$), thus giving rise to fully polarization insensitive integrated photonic circuits \cite{Corr18}.

Concerning the waveguide writing in transparent dielectric crystals, the most commonly adopted approach is that of inscribing by laser irradiation two parallel tracks of damaged and amorphized material, separated by a distance from 10 $\mu$m to 30 $\mu$m, which form the waveguide cladding. Waveguides fabricated in this way are called type II waveguides, and have been demonstrated for a great variety of crystals, including many rare earth doped matrices, several nonlinear crystals and diamond \cite{chen14}. More complex cladding-irradiation strategies have also been explored, with the aim of improving the vertical confinement of the light, or for creating photonic band-gap guiding structures (type III waveguides) \cite{rodenas19}. Finally, another possibility for fabricating channel waveguides in crystals by FLM consists in fabricating two parallel ablation trenches at the crystal surface for producing a guiding region between them (Type IV waveguide) \cite{chen14}. Instead, the creation of a positive refractive index change in crystals, usually associated to a local material densification, is very hard to achieve with direct laser irradiation. Therefore, the demonstration of type I waveguides has been successful only in a limited number of cases, e.g. lithium nibate \cite{atzeni18}, yttrium orthosilicate \cite{Seri18} and potassium dihydrogen phosphate \cite{huang15}.

Crystal waveguides fabricated by FLM typically show worse optical properties than what attainable in glass, suffering from higher propagation losses (> 1 dB/cm) and lower light confinement, with guided-mode field diameters ranging from 10 $\mu$m to 20 $\mu$m. This holds especially true in the case of type II-IV structures, preventing the use of these kinds of waveguides in curved geometries. Nevertheless, the possibility of writing optical waveguides in crystals remains of very high scientific interest, including for integrated quantum photonics applications, because they allow to take advantage of the peculiar physical properties that many transparent dielectric crystals offer.

\subsection{3D microstructuring}

Femtosecond laser pulses are capable of machining the surface of the dielectric, transparent substrate with ablation processes, which are driven by the same nonlinear absorption exploited to inscribe waveguide tracks in the bulk. By focusing the laser beam on the top surface of the substrate it is possible to drill precise holes, or excavate microtrenches. If low pulse-repetition rates are adopted, the ablation process is mainly non-thermal and may yield high resolution and no damage of the surrounding region.  However, during the ablation process debris can deposit close to or just upon the ablated region, which makes it difficult to realize clean microstructures. In addition, both the presence of the debris and the presence of the ablated structure itself modify the focusing conditions while the ablation proceeds deeper: this limits the depth and the achievable complexity of the realized structures.

An alternative approach is to perform laser ablation with the substrate  immersed in water, or simply with water wetting the surface to be machined, and starting from the bottom surface instead of the top one \cite{Li01}. In this way, while the ablation process proceeds towards the middle of the substrate, the laser beam always propagates through the clean substrate volume, without being affected by the microstructure that is being formed. In addition, the presence of water facilitates debris removal and makes it possible to excavate clean microchannels, where water can flow. This water-assisted laser ablation technique has enabled to fabricate microfluidic channels with an articulated path, drawn in the three-dimensions inside the substrate \cite{Hwang04}. Hollow chambers connected to the surface only by means of small passages \cite{An06}, or even microthreads in glass \cite{Dega17} have also been reported. 

A water-immersion objective can be used conveniently to perform, in the same session, both the inscription of waveguides and the fabrication of hollow microstructures by ablation. Only the irradiation parameters, such as pulse energy and translation speed, need to be changed between one kind of processing and the other, without the need of realigning the sample. Waveguides and microstructures are in this way perfectly positioned one relative to the other. Using this method, mechanical microcantilevers have been machined in borosilicate glass, with embedded optical waveguides, to realize optical-intensity modulators and switches working at a frequency of tens of kilohertz \cite{Spagn20}. In addition, deep microtrenches and suspended glass structures have been incorporated in reconfigurable waveguide circuits, controlled by thermo-optic phase shifters (see also the discussion in Section 5.2). In this case, the hollow microstructures guide the heat diffusion processes inside the substrate and enable a highly efficient thermal control of the circuit \cite{Cecca20} (see Fig.~\ref{fig:figureFLM}d,e).

A further, widely adopted method to perform femtosecond laser-assisted glass microstructuring is the so-called FLICE  technique (femtosecond laser irradiation followed by chemical etching), which exploits the enhanced sensitivity of laser-irradiated regions to chemical etching, in certain materials \cite{Marci01}. According to this method, tracks are irradiated in the bulk glass, terminating at one of the facets of the sample. The glass sample is later immersed in a chemical etching solution, such as an aqueous solution of hydrofluoric acid or potassium hydroxide. The etching agent preferentially attacks the irradiated region, giving rise to hollow microstructures along the irradiated lines \cite{Sugi05, Hnato06}.

As mentioned, this technique does not work well in all kinds of substrate, but preeminently in specific materials such as fused silica \cite{Marci01, Hnato06}. or photosensitive glass \cite{Sugi05} (e.g. Schott Foturan). In these substrates the high etching selectivity enables to machine three-dimensional microstructures with complex shapes. In particular, femtosecond-laser written waveguides with reasonably low losses can also be  fabricated in fused silica. It becomes thus possible to realize, in the same irradiation step and with a simple adjustment of the irradiation parameters, both optical waveguides and tracks that will be etched in a subsequent process, forming microfluidic channels that will be intrinsically aligned with the waveguides \cite{Osell07}. This technique have allowed to demonstrate optofluidic devices for sensing or other biological applications\cite{Bellini10,Crespi10,Bragheri12, Crespi12}, and even to integrate fluidic functionalities within the cladding of silica optical fibers \cite{Haque14}.

\subsection{Two-photon polymerization}

In the previous Sections we have seen how FLM can modify the properties of a transparent material (particularly its refractive index) or can selectively remove arbitrary portions of the material. Here we will see that FLM can also be used for additive manufacturing tasks through a process  known as two-photon polymerization (2PP) \cite{lee06}. The fundamental physical mechanism is again based on the exploitation of focused ultrashort pulses to trigger a nonlinear absorption process. However, this time the nonlinear absorption is performed inside a negative tone photoresist that absorbs in the UV. Using a near-infrared femtosecond laser source, a two-photon absorption process  triggers a local photopolymerization of the resist. Moving the focus inside the resist it is thus possible to polymerize 3D structures of arbitrary shape (see Fig.~\ref{fig:figureFLM}f,g), while the unirradiated resin is removed in a subsequent developing step. The nonlinear dependence of the absorption process on the intensity profile, as well as the fact that photopolymerization happens only above a given threshold of absorbed energy density, make it possible to achieve sub-diffraction-limit feature sizes with a careful tuning of the laser fluence \cite{tan07}. In addition, it is a maskless process that can be easily combined with preliminary optical inspection of the sample, enabling the possibility of in-situ lithography of photonic structures in specific locations where, e.g., single photon emitters have been found and characterized \cite{shi16}. \REV{Finally, although 2PP is applied to photosensitive resins, while the previous micromachining processes regard glasses and crystals, all of them can be implemented on the same femtosecond laser micromachining equipment, by carefully designing the process workflow \cite{Wu2014}.}

\subsection{Advantages of FLM for integrated quantum photonics}

It is evident from the previous discussion that FLM has many unique features with respect to all the other integrated photonic platforms. Such features have clear advantages also for quantum photonic applications. 

Let's first discuss the versatility of the process. Not only it encompasses transformative, subtractive and additive technologies in the same tool, but those technologies can also be easily combined to produce complex 3D quantum devices with a very limited number of processing steps as compared to photolithographic technologies.  In addition, the versatility also regards the different materials that can be processed with FLM. It has been shown that almost all materials have a suitable processing window that allow the waveguide writing process. This flips the usual way of thinking in designing a photonic component, i.e. from how to integrate a given functionality in the material of the integrated photonic platform, to finding the processing window to produce an integrated photonic circuit in the best material for that functionality. Naturally, this approach has a perfect match with the modular architecture of the quantum system described in Section 2.1, where one can envisage the optical interconnection of separately optimized components without the constraint of a common substrate material. 

Focusing more specifically on the properties of the laser written photonic circuits as compared to the other platforms, we should first emphasize that FLM has the unique possibility of writing 3D photonic circuits, while the other technological platforms are mainly confined to 2D layouts. This feature can be exploited in devising more compact devices or in removing the usual limitation of first-neighbour interaction in designing quantum protocols on a photonic chip. 

Laser-written waveguides have mode sizes and refractive indices very close to those of optical fibers, therefore extremely low-loss edge coupling can be easily achieved. This is an important feature that further supports a modular architecture. Not only coupling losses are low, but also propagation losses are typically around 0.1 dB/cm or better. The downside of the similarity of the waveguide confinement to that of an optical fiber is the limited minimum radius of curvature that can be implemented without additional losses. This brings the miniaturization capability of FLM circuits at the level of the silica-on-silicon platform, but significantly worse than most of the alternative ones. Another important advantage of the FLM platform is the low birefringence of the written waveguides and the possibility to produce polarization insensitive couplers. This allows the manipulation of the quantum information encoded in the photons irrespectively of their polarization state, while other platforms require a fixed polarization for proper operation. This is a great simplification in the global system, in particular in the modular architecture, where it may be complicated to preserve the polarization state in the optical interconnections. 

Finally, all the versatility that we have emphasized has a cost in terms of mass-production capability. In fact, FLM is a serial writing technique and cannot compete with parallel photolithographic microfabrication for very large volume of devices. The gap is progressively reducing thanks to the much smaller number of steps required to produce a complete photonic circuit, to the increasing processing speeds that have now reached  several cm/s, and to the various strategies that are being developed to parallelize the irradiation process \REV{ \cite{Pospiech2009,Mauclair2009,Zandrini2019,Manousidaki2020}}. In addition, mass production is not yet an issue in quantum technologies, since potentially commercial devices are still under development, whereas the capability of providing custom-tailored components that can be prototyped rapidly and at low cost is a very relevant feature.

In the following we will discuss several examples where all these advantages have been exploited.

\section{Single photon sources}
Nonclassical states of light are the core quantum resource that lie at the heart of quantum photonic technologies. Most of current experiments involving FLM integrated photonic circuits rely on external single photon sources of various kinds, from spontaneous parametric down-conversion sources \cite{Crespi13a} to quantum dots \cite{anton19} or defect-based \cite{white20} sources. This is made possible by the excellent connectivity of laser-written devices with standard optical fibers. However, the direct on-chip generation of quantum states of light is a very important functionality, subject of extensive research activity by several groups for a twofold motivation: on the one hand, generating quantum light directly inside the photonic circuit allows eliminating the coupling losses that unavoidably occur when interfacing a device with an external source; on the other hand, waveguide-based single photon sources permit quantum state engineering with multiplexed parallel sources in an interferometrically stable and scalable fashion, for creating complex quantum states of light which are out of reach for bulk setups.

Integrated single photon sources can be classified in two distinct categories \REV{\cite{eisa11}}: i)~probabilistic sources, where a bright pump beam is coupled to a waveguide fabricated in a nonlinear medium,  for the generation of photon pairs either by spontaneous parametric down conversion (SPDC) or spontaneous four wave mixing (SFWM) processes; ii)~deterministic sources, which exploit the fluorescence of single quantum emitters, e.g. individual quantum dots or point defects in crystals. FLM contributed to the development of both types of integrated sources, by providing the necessary circuitry for manipulating the pump light and the generated photons, by fabricating the nonlinear waveguides, and also by creating the emitting defects with a deterministic control of their position in the host matrix.

\begin{figure*}
\centering
\includegraphics[width=\linewidth]{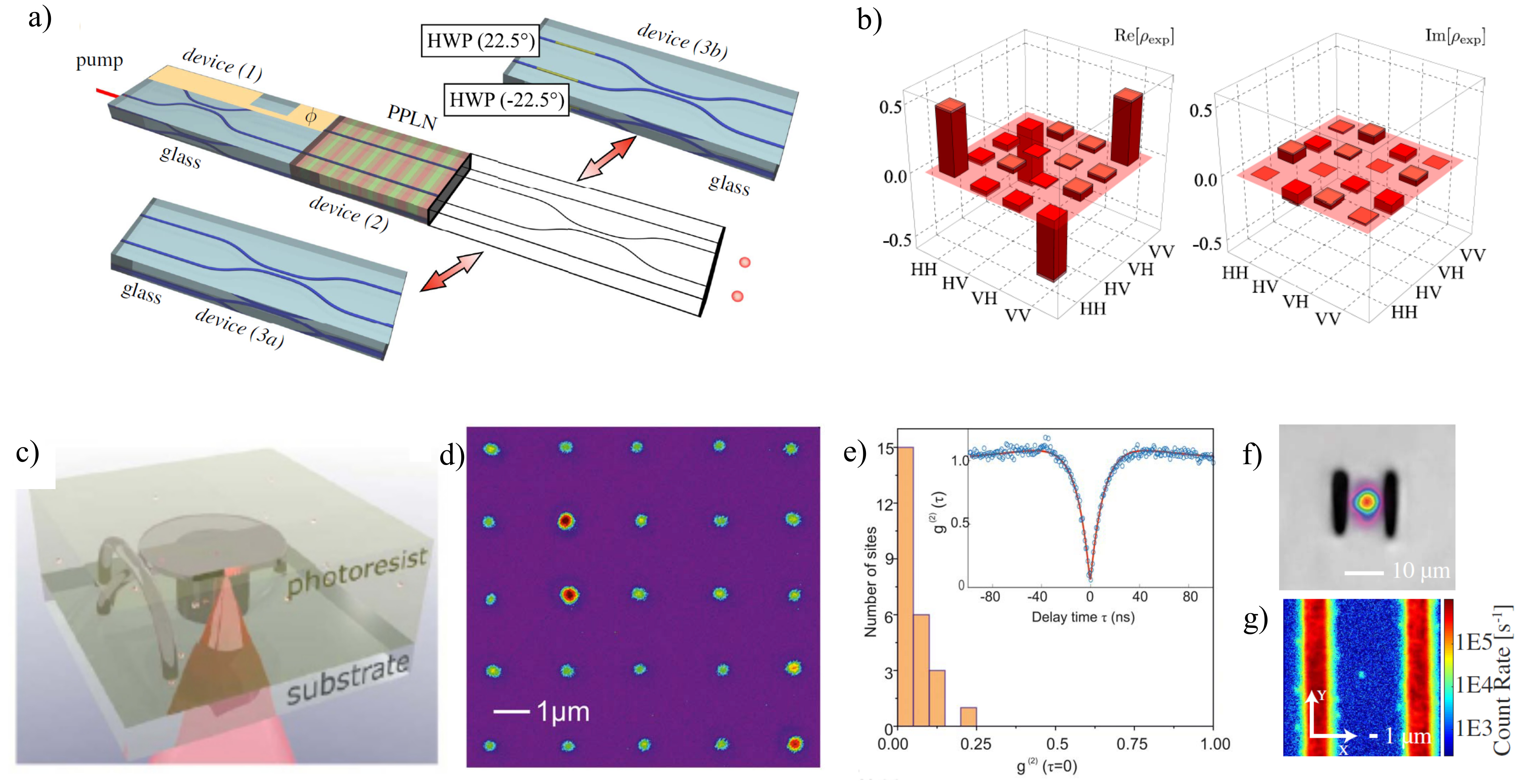}
\caption{\label{fig:sources} (a)~Schematic of the integrated single photon source presented in \cite{atzeni18}. A first reconfigurable glass chip is used for pump splitting. The PPLN chip is used for photon pairs generation. The third stage can be interchanged between device 3a and 3b in order to engineer different type of biphoton states. 
\REV{(b) Results of the quantum state tomography performed
on the biphoton states produced by the device presented in \cite{atzeni18}, when set to generate the polarization-entangled $|\phi^{-}\rangle$ state. Reported fidelity with the theoretical state is 93 \%.}
(c)~Schematic of the 2-photon polymerization process employed in \cite{schell2013} for fabricating integrated nanophotonic devices with embedded nanodiamonds. (d)~Fluorescence map obtained from a 5~x~5 square lattice of NV centers fabricated by FLM in \cite{chen19}. Lattice parameter is 2 $\mu$m. (e)~Histogram showing the second order autocorrelation results obtained for the fluorescence light of all NV centers shown in panel (d). In the inset, a whole $g^{(2)}(t)$ trace measured for a single defect is shown. (f)~Transverse cross section and near field image of the guided mode of the type II waveguide fabricated in diamond and presented in \cite{hadden18}. (g)~Photoluminescence image of the longitudinal profile of the same waveguide as in panel (f), where it can be clearly noticed the presence of an individual NV center localized within the core. The panels of this figure contain pictures adapted from Refs. \cite{atzeni18}, \cite{schell2013}, \cite{chen19} and \cite{hadden18}.}
\end{figure*}

\subsection{Photon generation based on nonlinear phenomena in waveguides}

Probabilistic sources are relatively easy to implement. However, they suffer from the drawback of producing photons randomly distributed in time. This limits their employment for creating complex quantum states involving a large number of photons. Nevertheless, the possibility of exploiting heralding or post-selection mechanisms makes this kind of sources highly appealing for applications. In addition, having the pump beam tightly confined over the whole waveguide length helps in increasing the overall efficiency of the generation process, thus resulting in enhanced brightness of the source \cite{tanzilli01}.

The first demonstration of heralded single photon generation in a FLM-written waveguide is due to Spring et al. in 2013 \cite{spring13Source}. In this experiment the authors fabricated a 4~cm long waveguide in fused silica, and exploited the weak glass $\chi^{(3)}$ non-linearity for generating non-degenerate photon pairs by SFWM, with the signal and idler photons at the wavelengths of 676 nm and 790 nm respectively. Notably, the waveguide was fabricated by shaping the writing beam through adaptive optics elements in order to increase the ellipticity of the waveguide cross section and to enhance the value of waveguide birefringence up to $10^{-4}$. This allowed to phase-match the SFWM process at wavelengths where the photonic noise generated by spontaneous Raman scattering of the pump in the silica substrate becomes negligible. The measured purity of the quantum states reached 0.86 without performing spectral filtering at the waveguide output, and this value was limited by residual birefringence fluctuations along the waveguide, estimated as $\delta(\Delta n)<3\cdot10^{-6}$.
In 2014, a different integrated single photon source, based on photonic circuits written in borosilicate glass, was proposed for the generation of spatially multiplexed, heralded single photons by SPDC \cite{meany14}. In this case, the nonlinear stage was made with a periodically poled lithium niobate (PPLN) chip, fabricated by soft proton exchange, containing four identical waveguides. A laser-written 1~x~4 beam divider was directly butt-coupled to the PPLN chip input, in order to pump coherently the four PPLN waveguides with a single laser source. The generation process was phase matched for producing non-degenerate photon pairs at the wavelengths of 1550 nm (signal) and 1312 nm (idler). A second FLM-written photonic chip was cascaded after the generation stage, containing four dichroic directional couplers, capable of separating the signal and the idler photons in distinct spatial modes with an average extinction ratio of 10 dB. The chip was then connected to an external filtering stage, in order to suppress the pump light and for increasing the photons purity. Finally, a bulk system of fast optical switches was used to route the generated photons to a single spatial mode, and to demonstrate a fourfold enhancement of the generation rate with respect to a single source.
A similar hybrid approach, where a lithographic PPLN chip is combined with FLM-written borosilicate circuits, was used in a subsequent experiment in 2016, for the demonstration of an integrated source of indistinguishable photon pairs in the telecom C-band \cite{verg16}. A first glass chip was employed to split the pump light and route it to two PPLN waveguides, for generating pairs of non-degenerate photons. A second glass chip was cascaded afterwards, containing the required elements for separating the signal (1560 nm) and the idler (1310 nm) photons, and for engineering the quantum state of the signal photon pairs. In particular, a thermally reconfigurable Mach-Zehnder interferometer was used as a variable beam splitter, in order \REV{to} combine the signal photons, and produce either a separable state $\psi=|1\rangle_1|1\rangle_2$ or an entangled state in the photon number basis of the form $\psi\propto|2\rangle_1|0\rangle_2-|0\rangle_1|2\rangle_2$ (a two-photon NOON state) depending on the Mach-Zehnder setting and by post-selecting the cases where the photons are generated simultaneously in the two PPLN waveguides.
Finally, in another experiment from 2018, a hybrid glass-PPLN-glass device was used for the on-chip generation of different quantum states of light, based on the creation of degenerate pairs of photons at 1560 nm by SPDC \cite{atzeni18}. A modular approach was adopted, where the second glass chip after the PPLN waveguides could be exchanged, in order to engineer different two-photon quantum states, ranging from product and NOON states, to polarization-entangled pairs (see Fig.~\ref{fig:sources}a\REV{,b}). In this case, all components have been entirely fabricated by FLM. In particular, two identical type I waveguides have been written within the bulk of a z-cut PPLN sample at the depth of 100 $\mu$m by employing a multiscan approach. This allowed to finely control the waveguide cross section and obtain single mode guidance at both the pump (780 nm) and photons wavelengths. 
\REV{In addition, the two waveguides exhibited identical spectral behavior, as measured with second harmonic generation experiments performed with classical light.}

\subsection{Photon generation based on induced defects in diamond}

Among the different types of individual quantum emitters currently studied as deterministic sources of quantum light, single negatively-charged Nitrogen-Vacancy (NV) centers, i.e. optically active point defects that form in diamond when a substitutional nitrogen atom bounds together with a carbon vacancy, are particularly appealing. In fact, they show a very high room-temperature radiative quantum efficiency upon optical excitation, as well as a short decay time of the excited state. The antibunched nature of single NV center fluorescence has been widely demonstrated \cite{kurts00}, and this makes this system a very attractive candidate to realize deterministic single photon sources.

The first attempt of integrating an NV center source within a laser-written device is due to Schell et al. in 2013 \cite{schell2013}. In particular, the authors fabricated a set of 3D arch-shaped waveguides and microdisc resonators using the two-photon polymerization technique (see Fig.~\ref{fig:sources}c), employing an acrylate polymer photoresist (PETTA) mixed with an ethanol-based nanodiamonds suspension (mean diamond diameter of 25 nm). Such nanodiamonds naturally contained NV centers, which remained photostable even after the laser irradiation for the FLM process, and resulted dispersed within the volume of the photonic devices. A confocal microscope setup was then used for exciting individual defects. The analysis of the fluorescence emission, collected at the waveguides output and fed into a Hanbury-Brown-Twiss interferometer, clearly confirmed its single-photon nature. In a subsequent experiment, the same authors showed that this approach is compatible also with a waveguide-based excitation of the NV centers \cite{shi16}. In this case, a peculiar 3D crossed-waveguide coupler configuration was employed, in order to separate the excitation mode from the collection mode, and suppress the photonic noise created in the waveguide by Raman scattering and from fluorescence of the polymerized photoresist. Remarkably, in this experiment the photonic devices have been fabricated around pre-characterized defects, selecting those with best quantum emission features.

FLM can be used also for the direct writing of NV centers at the surface \cite{kono17} and inside the volume of bulk diamond \cite{liu13, pimenov16, chen17, hadden18, chen19}. In particular, the first demonstration of this capability, together with the accurate characterization of the single photon nature of the fluorescence emitted by the fabricated defects, is due to Chen and co-workers \cite{chen17} in 2017. They employed single pulse irradiation followed by a 1000~$^{\circ}$C thermal annealing (few hours duration) of the whole sample for creating single NV centers per irradiated point with $\approx$~45~\% yield. This process allowed inscribing high-quality single defects with an in-depth positioning accuracy of 700 nm and a transverse positioning accuracy of 200~nm. In a subsequent work \cite{chen19}, this inscription method has been improved by substituting the thermal annealing process with a second laser irradiation step, in order to heat locally the diamond substrate at the NV center location. In addition, the FLM setup was modified by adding an in-line confocal microscope, exploiting the fabrication microscope objective also for collecting in real time the fluorescence of the defects during the heating step, thus providing an online feedback for tailoring the irradiation time for obtaining high quality single NV centers. In this way, the authors significantly improved the defect fabrication yield, up to 96~\%, and the positioning accuracy (33~nm in-plan accuracy and 200~nm in-depth accuracy). In Fig.~\ref{fig:sources}c the fluorescence traces from a $5 \times 5$ matrix of NV centers fabricated in this way are shown. Second order autocorrelation measurements resulted in $g^{(2)}(0)<0.2$ (see Fig.~\ref{fig:sources}e) for all of them, confirming the antibunched nature of the emission. In addition, the authors have shown that such defects can be arranged in large 3D matrices, containing up to 2000 NV centers. Beside quantum light generation, such defects can also be employed as long lived spin qubits, with a spin excitation coherence time greater than 500 $\mu$s \cite{step19}. The direct FLM writing of optically active single defects is a particularly interesting method, since it has been demonstrated that it can be applied to other materials besides diamond, including silicon carbide \cite{castel18}, cubic boron nitride \cite{buividas15} and gallium nitride \cite{saleem18}.

Finally, ultrafast laser writing is also suitable for the inscription of buried type-II and type-III optical waveguides in diamond \cite{courvo16, sotillo16} (see Fig.~\ref{fig:sources}f). Interestingly, in 2018 Hadden et al. \cite{hadden18} have shown that waveguides and NV centers writing by FLM are two complementary techniques that can be combined for the  inscription of the NV centers directly coupled with the waveguide mode. In particular, they first fabricated a type II optical waveguide in the bulk of a diamond sample, and then inscribed a set of NV centers, employing the same procedure presented in \cite{chen17}, precisely localized within the waveguide core (see Fig.~\ref{fig:sources}g). It was then shown that the thermal annealing step did not alter the waveguide properties, and that the waveguide could be used both for exciting the NV center and for collecting its fluorescence emission. These results proved that FLM has a significant potential in developing diamond-based integrated quantum photonics platforms.

\section{On-chip manipulation of quantum states of light}

Integrated photonic circuits are made of straight or bent waveguide segments, directional couplers and integrated waveplates. Combinations of these elements indeed allow to perform a unitary transformation on the state of the injected photons (neglecting photon losses or post-selecting output states with the same number of photons as the input).

From a physical point of view, the operation of these devices is based on a combination of classical and quantum interference of photons. A striking advantage of the adoption of an integrated platform lies in its inherent interferometric stability, which together with the miniaturization properties, makes it possible to realize optical circuits that would be too large to assemble, or very hard to operate, with bulk components on an optical table. 

In the following we will review the main achievements in the on-chip manipulation of quantum states, which have been accomplished by FLM waveguide circuits. We will broadly distinguish between static and dynamically-reconfigurable photonic circuits. The former cannot actively modify their function, which is thus fixed upon fabrication. The latter ones, on the other hand, contain components whose operation is governed by external signals (e.g. electrical currents), and can actively change in time the performed unitary transformation.

\subsection{Static photonic circuits}

Manipulation of path-encoded photonic qubits, which essentially requires to interfere identical photons on many different paths, comes natural in the integrated setting \cite{Politi09}, and is indeed perfectly possible with femtosecond-laser-written devices. Shortly after the first demonstration of the operation of a silica-on-silicon optical chip in the quantum regime \cite{Politi08}, it was shown that quantum interference could be reliably observed also in femtosecond-laser written waveguide circuits \cite{Marsh09}. 

Static circuits, written by femtosecond lasers, have proved capable to prepare and characterize high-order single-photon entangled states on many paths \cite{Grafe14}, to produce two-photon path-encoded Bell-states \cite{Li20}, or to efficiently perform quantum state tomography of single and multi-photon states involving several optical modes \cite{Titch18}. 
By interfacing a waveguide Mach-Zehnder interferometer with an etched microchannel, in a single optical chip fully fabricated by femtosecond laser pulses, it was possible to perform two-photon quantum interferometry and experimentally retrieve the concentration of biological solutions \cite{Crespi12}.

Special multi-port interferometers \cite{Spagn13t, Spagn13b, Tillm15, Crespi16, Viggia18njp, Viggia18sb,munz21} have allowed to investigate fundamental aspects of multi-photon quantum interference. In particular, symmetric multi-ports \cite{Crespi16,Viggia18njp, Viggia18sb,munz21} were employed to demonstrate experimentally massive effects of quantum destructive interference, which suppress many of the possible output states; these integrated devices could also find application in assessing the quality of multi-photon sources.

\begin{figure*}
\centering
\includegraphics[width=\linewidth]{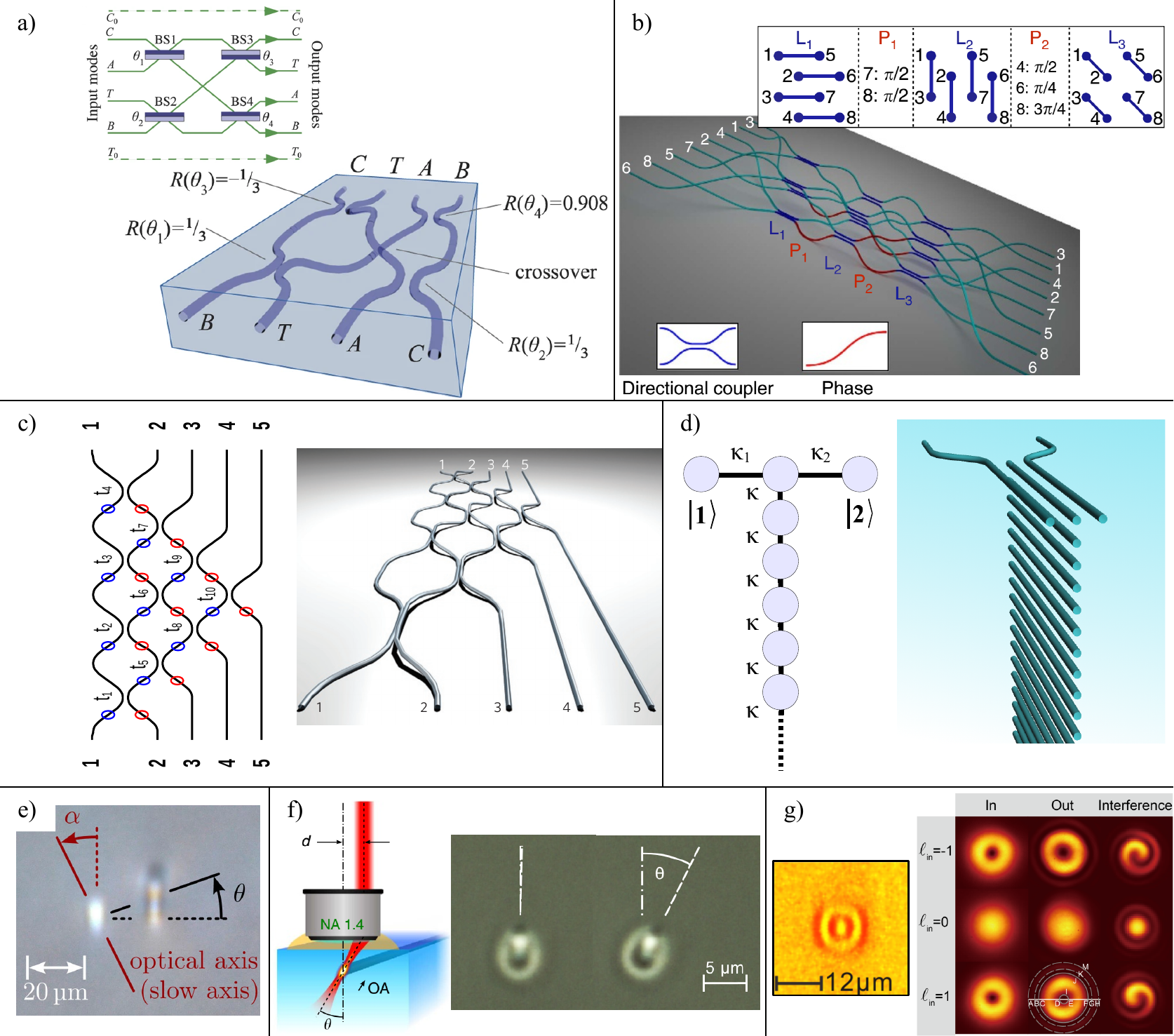}
\caption{\label{fig:static}Examples of static, femtosecond-laser-written photonic circuits adopted for quantum applications. (a) Heralded controlled-Z gate for path-encoded photonic qubits: the bulk-components schematic is depicted on the top-left; on the right a pictorial representation of the femtosecond-laser written circuit. The optical waveguides pass one over the other without crossing, at the center of the circuit. (b) Three-dimensional circuit implementing the quantum Fourier transform of eight optical modes. The circuit is composed of three sections and coupling between modes in each step ($L_i$) are shown in the inset. The implemented phase shifts in each step ($P_i$) are also indicated. (c) Concept scheme and three-dimensional representation of a 5-mode circuit, realized according to the Reck's layout, which implements an arbitrary unitary transformation between the input and output modes. (d) Example of circuit for analogical quantum simulation: the quantum system represented on the left, consisting of two discrete sites ($\left|1\right\rangle$ and $\left|2\right\rangle$) coupled to a linear chain of sites, can be simulated by the photonic circuit depicted on the right. (e) Cross-section of an integrated waveplate (microscope picture): the birefringence axis of the waveguide \REV{(on the} left) can be rotated by writing an additional track (on the right) which induces stress on the former. (f) An alternative approach to tilt the birefringence axis of a waveguide of an angle $\theta$ (see cross-sections on the right) is to employ, for the laser inscription, a large-numerical aperture objective: the writing laser beam, which must not fill the objective completely, impinges on the objective entrance at a distance $d \propto \tan \theta$ from the optical axis. (g) A ring-core waveguide is inscribed by juxtaposing irradiated tracks, to describe a circle. A straight waveguide segment with such shape is able to propagate optical modes with non-vanishing values $l_m$ of orbital angular momentum. Pictures in the panels have been adapted respectively from Refs.~\cite{Meany16}, \cite{Crespi16}, \cite{Crespi13a}, \cite{Crespi15}, \cite{Heil14}, \cite{Corr14} and \cite{Chen18}.}
\end{figure*}

It is worth noting that, in an integrated platform, continuously-coupled waveguide arrays allow to implement interactions among optical modes that have no analogous in the bulk realm \cite{Spagn13t, Grafe14, Titch18}. In addition, the FLM technology yields the unique capability of inscribing circuits with three-dimensional layouts, which further enhances stability and compactness. In fact, waveguides can pass one over the other, if needed, without intersecting \cite{Meany16} (see Fig.~\ref{fig:static}a), and circuit architectures, impossible to realize with a planar technology, are made feasible \cite{Grafe14, Crespi16,Viggia18njp, Viggia18sb}. Interestingly, special three-dimensional arrangements of directional couplers have been reported, inspired to the Cooley-Tukey Fast-Fourier-Transform algorithm \cite{Cooley65,Barak07,Crespi16,Viggia18njp, Viggia18sb} (see Fig.~\ref{fig:static}b): such architecture can be used to realize a significant class of linear optical networks with a substantial reduction in the number of needed components, and notable resilience to losses and fabrication tolerances \cite{Flam17}.

In addition, static multi-port interferometers, fabricated by femtosecond laser pulses, have played a significant role in the recent years as hardware for Boson Sampling experiments. Boson Sampling, a kind of experiment conceived by Aaronson and Arkhipov in 2011 \cite{Aaron11}, consists in sampling from the output distribution of multiple interfering bosons, evolved along a random unitary transformation. Such task is of interest because it is deemed intractable for classical computers as the system size scales up; thus, it can provide a route to demonstrate a quantum experiment that is hard to simulate with classical resources, showcasing a kind of quantum advantage \cite{Zhong20}. In the photonic version, Boson Sampling consists in sampling the output distribution of multiple photons that have propagated in a large random interferometer \cite{Brod19}. 

FLM indeed provided the photonic hardware (in detail, five-mode integrated interferometers) for two \cite{Tillm13, Crespi13b} of the four \cite{Tillm13, Crespi13b, Broome13, Spring13} first-time demonstrations of Boson Sampling experiments in 2013. Following experiments, which investigated different aspects of the Boson Sampling problem, harnessed fetmosecond-laser-written interferometers with up to 13 optical modes \cite{Spagn14,Benti15,Giorda18,Agre19}. Note that an interferometer that realizes a random arbitrary unitary transformation may not be  trivial to design in practice. Theoretical schemes are known, by Reck \cite{Reck94} and Clements \cite{Clem16}, which implement an arbitrary unitary by cascading directional couplers with different power-splitting ratios, and different phase shifts in between them. However, in a planar network of waveguide components, changes in the coupler geometry (to modulate the splitting ratio) will produce changes also to the length of the interconnections, thus modulating also the phase shifts, making it difficult to control independently all the parameters. Of note, Ref.~\cite{Crespi13b} demonstrated a novel design technique for the optical circuit to overcome this problem, by exploiting the unique three-dimensionality given by femtosecond laser waveguide inscription (see Fig.~\ref{fig:static}c). Different phase shifts are produced in the circuit by properly deforming the S-bends that connects the cascaded couplers. Instead, the splitting ratios are modulated by rotating one arm of the directional coupler out of the plane, thus making the two waveguides closer or farther in the interaction region without changing the length of the connecting waveguides. 

In all the experiments described above, photons have fixed polarization states, and the only degree of freedom that the integrated circuit is manipulating is in which spatial mode the photon is propagating. Actually, femtosecond-laser-written waveguides conceal a more peculiar potential for the manipulation of photons in multiple degrees of freedom. Shortly after the first demonstration of a quantum application of a femtosecond laser written circuit \cite{Marsh09}, Sansoni et al. reported \cite{Sans10} that a femtosecond-laser-written directional coupler in borosilicate glass could safely interfere also polarization entangled photons. Ciampini et al. \cite{Ciamp16}, using a pair of analogous directional couplers, demonstrated that hyperentangled states of two photons could be successfully manipulated (i.e. photons entangled both in the polarization and path degrees of freedom). Recently, Chen et al. \cite{Chen18} were even able to propagate single-photon states with orbital angular momentum in specially tailored waveguides, again inscribed in borosilicate glass by juxtaposing irradiated tracks to form a ring-shaped core (Fig.~\ref{fig:static}g).

The capability of supporting and manipulating the polarization degrees of freedom make femtosecond-laser-written circuits suitable for quantum protocols based on polarization encoding. On the one hand, by using specific geometries for the directional couplers \cite{Sans12} or special irradiation recipes ensuring ultra-low birefringence \cite{Corr18}, polarization-transparent circuits can be produced \cite{Sans12, Crespi13a, Chap16, Pitsios17p, Zeuner21}. On the other hand, by harnessing waveguide birefringence, partially-polarizing \cite{Crespi11} and polarizing \cite{Ferna11, Pitsios17g, Zeuner18} integrated beam splitters (directional couplers) can be inscribed, enabling the realization of integrated two-qubit CNOT gates either working in post-selection \cite{Crespi11} or with heralding  \cite{Zeuner18}. In addition, the birefringence axis of the waveguide can be locally tilted to produce integrated waveplates, either by juxtaposing laser-irradiated tracks that produce stress oriented at the desired angle in the substrate \cite{Heil14} (Fig.~\ref{fig:static}e) or by employing asymmetric focusing of the writing laser beam \cite{Corr14} (Fig.~\ref{fig:static}f). The latter components can implement arbitrary one-qubit gates \cite{Heil14} and, together with the above-mentioned two-qubit ones, constitute an in-principle complete toolbox for integrated quantum computing with polarization-encoded qubits. Further devices based on quantum information encoded in the photon polarization are an integrated QKD transmitter \cite{Vest14}, demonstrated by exploiting a polarization insensitive network of directional couplers, and an integrated apparatus for polarization full state tomography of two-qubit states, realized by combining integrated waveplates and polarization insensitive couplers on chip \cite{Corr14}.

A different set of applications of static photonic circuits are disclosed by the observation that\REV{, within the paraxial approximation \cite{lax75}, light propagation in dielectric structures is described by a Schroedinger-type equation \cite{Marte97, Longhi09,Szameit10}. In this picture, an array of femtosecond-laser written waveguides implement a lattice of quantum sites, whose mutual coupling is described by a tight-binding model \cite{Chen21}, and time evolution is mapped onto the propagation coordinate. In addition, proper bending of the waveguides implement effective external forces \cite{Longhi09, Corr13, Rechtsm13}. In the realm of quantum optics,
the optical modes of a set of waveguides, populated by single- or multi-photon states, represent effectively discrete quantum states populated by bosonic particles. 
By tailoring the array geometry and the optical properties of the waveguides, it is indeed possible to implement diverse Hamiltonians and, by injecting single or multi-photon light states, perform quantum simulation of different physical systems. In fact, b}y this method, quantum walks of one \cite{Tang18e,Tang18h, Shi20} and two \cite{Owens11, Sans12, Crespi13a, DiGiu13, Poulios14} particles were demonstrated on different kind of graphs. \REV{Furthermore, t}his setting allows for fundamental studies, such as investigating the influence of noise on quantum coherence \cite{Perez18}, probing dynamics due to PT symmetry \cite{Klauck19}, demonstrating the properties of photonic topological insulators \cite{Tamb18, Wang19t, Wang19d, Klauck21}. As a matter of fact, the range of simulatable phenomena is indeed wide and encompasses, among the examples present in the literature, also quantum decay dynamics \cite{Crespi15} (Fig.~\ref{fig:static}d), Bloch oscillations of NOON states \cite{Lebugle15}, perfect state transfer protocols on a spin chain \cite{Chap16}. 

We note that the three-dimensional capabilities of the FLM technique indeed allow to implement graph topologies that would not be possible in a planar circuit setting \cite{Owens11, Poulios14, Perez18, Crespi15}. In addition, if the circuit is polarization transparent, a polarization-entangled photon pair can be used, depending on the specific entangled state, to simulate the dynamics of anyonic or fermionic particles \cite{Sans12, Crespi13a, Crespi15}. 
A further advantage of this fabrication technology is that circuits implementing systems with slightly different parameters can be fabricated with rapid turnaround and in a cost-effective way \cite{Crespi13a, Crespi15, Perez18, Tang18e, Tang18h, Shi20}, thus allowing to conveniently explore different conditions or infer statistical quantities from the experiments. 

\subsection{Dynamically-reconfigurable photonic circuits}

Dynamic operation and reconfiguration of integrated waveguide circuits is typically achieved by externally controllable phase elements. Depending on the substrate, diverse physical effects may be exploited to provide localized and reversible changes of the refractive index of the waveguide structure, which in turn provide an additional phase delay to the propagating light. To date, active phase control in femtosecond-laser written waveguide circuits has been reported by exploiting the electro-optic \cite{Li06}, elasto-optic \cite{Humphr14}, or thermo-optic effect\cite{Chabo15,Flami15}. Indeed, such active phase elements are essential components in a complete quantum-photonics platform \cite{Politi09, Matthews09}, as even few dynamic phase shifters applied on a quantum waveguide circuit tremendously increase its potentials, by allowing to perform quantum protocols with parameters dynamically changing in time. 

The first quantum-optics experiment which included femtosecond-laser written waveguides with actively-modulated optical properties was reported by Humphreys et al. \cite{Humphr14}. In that work, the birefringence of an optical waveguide, inscribed at shallow depth in a fused silica substrate, was modulated by a mechanical pressure on the chip surface. The change in birefringence was visualized as a two-photon interference fringe exploiting a bulk polarization interferometer. Exploitation of the elasto-optic effect yields interesting potential in the quantum-photonics framework. In fact, mechanical piezo-electric actuators have low power consumption, which would make their use possible even in a cryostatic environment. Time response is dictated by the elastic properties of the material and can be in the order of microseconds or better \cite{Humphr14}. Furthermore, the applied mechanical stresses could be localized in a tight volume and cross-talks between neighbouring devices could be negligible.

\begin{figure}
    \centering
    \includegraphics[scale=0.95]{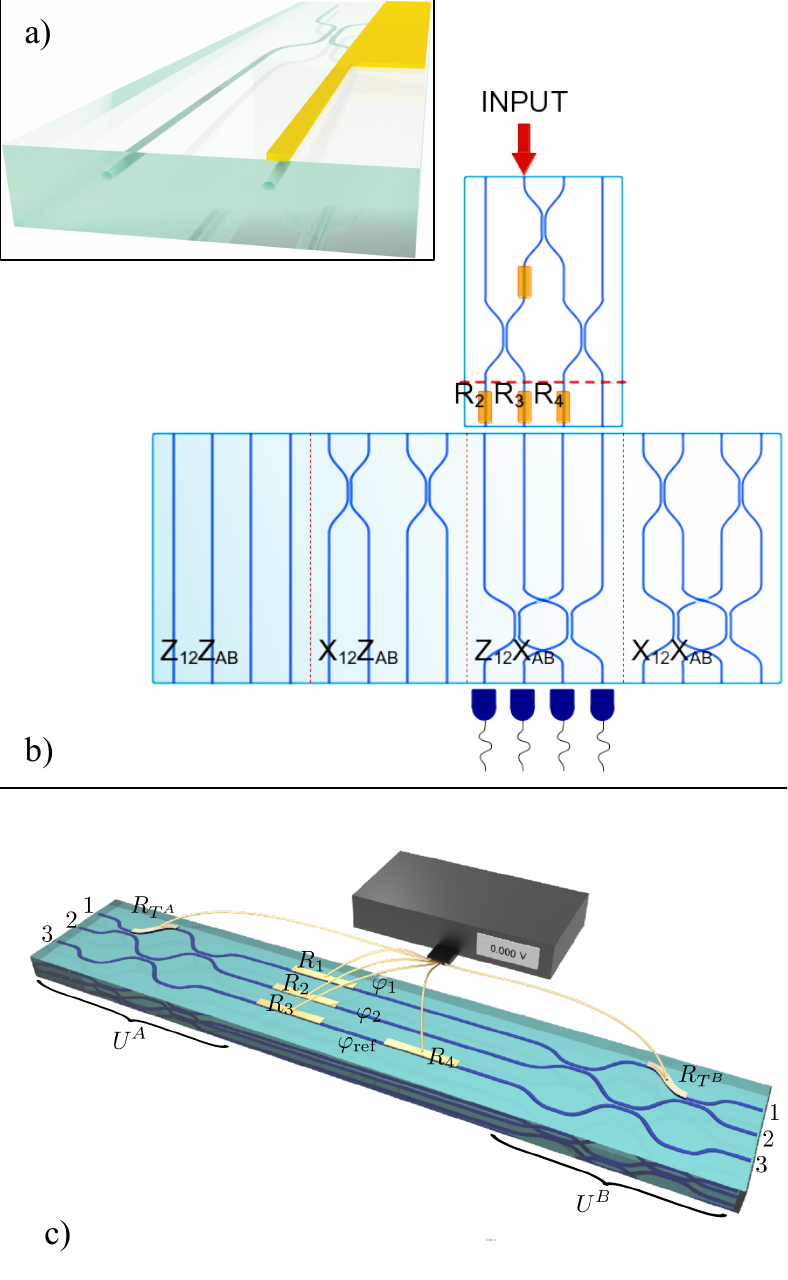}
    \caption{(a) 3D section of a reconfigurable Mach-Zehnder interferometer: a metallic heater is patterned on top of one of the interferometer's arms. (b) Pictorial scheme of the two cascaded photonic chips employed in Ref.~\cite{Crespi17}. The first one encodes a two-qubit path-encoded state of a single photon. The second chip, in conjunction with the four single-photon detectors at the output, enables to perform different measurements on the two-qubit state. Thermo-optic phase shifters, fabricated as metallic resistors on the first chip, allow to sweep through several different two-qubit states (R1) and to calibrate the phase terms at the interface (R2, R3, and R4). (c) Scheme of the reconfigurable three-arm interferometer reported in Ref.~\cite{Polino19}. Three straight waveguide segments are included between two multiport splitters, which implement unitary transformations $U_A$ and $U_B$. The device is equipped with six thermo-optic phase shifters,  which dynamically control phase terms inside the unitary transformations $U_{A,B}$ and in the interferometers arm. The panels of this figure contain pictures adapted respectively from Refs.~\cite{Cecca19}, \cite{Crespi17} and \cite{Polino19}.
    \label{fig:reconfigurable}}
\end{figure}

However, it is not trivial to integrate on a glass substrate mechanical micro-actuators with adequate features, both in terms of developed force and device miniaturization. As a matter of fact, to our knowledge and with the exception of Ref.~\cite{Humphr14}, all quantum photonics experiments to date, which involve reconfigurable femtosecond-laser written circuits, have exploited thermo-optic phase shifters. The latter devices are indeed much simpler to fabricate. A thermo-optic phase shifter is based on a metallic resistive heater, which is realized above a given waveguide segment, on the top surface of the optical chip (see Fig.~\ref{fig:reconfigurable}a). Heat dissipation on the resistor, by Joule effect, provokes a localized temperature increase in the substrate. This results in a localized refractive index variation, and thus a phase shift, which is typically proportional to the dissipated electrical power, and hence easily controllable in a dynamic way.

Thermo-optic phase shifters were demonstrated for the first time in femtosecond-laser written optical circuits in 2015, in two independent papers by Chaboyer et al. \cite{Chabo15} and Flamini et al. \cite{Flami15}. In the former \cite{Chabo15}, a three-dimensional three-arm interferometer was realized in a borosilicate glass substrate, and localized heating was provided by an array of alumina thick-film resistor mounted onto the chip. In the latter work \cite{Flami15}, Mach-Zehnder interferometers were laser-written in the glass substrate and, after waveguides inscription, a gold layer was deposited on top of the glass substrate. Resistors were patterned on such metallic layer using again femtosecond laser pulses, precisely aligned to the underlying waveguides. Both these works demonstrated circuit operation in the quantum regime, showing two-photon interference fringes as a function of the dissipated thermal power. 

Following these two first demonstrations, a few quantum experiments were reported, which exploited femtosecond-laser-written circuits with dynamic elements.

Crespi et al. \cite{Crespi17} demonstrated non-classical features of single photons (in detail, the single-particle quantum contextuality), using two cascaded optical chips (see Fig.~\ref{fig:reconfigurable}b). The first one contained a ramified circuit, equipped with four distinct thermal phase shifters realized as in Ref.~\cite{Flami15}. These phase shifters enabled the preparation of different states of a single photon delocalized on four optical paths (effectively encoding two qubits), and ensured phase adjustment at the chip interface. The second chip had static functioning and contained distinct waveguide circuits to perform different measurements on this prepared state, which were selected by properly aligning the preparation circuit in the first chip with the chosen measurement circuit in the second one.  The possibility to select a given circuit in the second chip, only by relative translation between the two, provided indeed another degree of reconfigurability to the experiment.

A similar cascaded-chips approach was employed also in Ref.~\cite{Pitsios17p} by Pitsios et al., to perform a quantum simulation experiment. Here the first circuit, operating statically, simulated the dynamics of a spin chain with five sites, exploiting the discrete-time quantum walk of two entangled photons. The second chip, featuring two thermo-optic phase shifters, allowed to interfere the output modes and characterize the output state in different bases (depending on the values of the actively-controlled phases). By such second chip it was thus possible to certify the presence of entanglement in the quantum state after the evolution in the first chip. 

Three-arm interferometers, equipped with up to eight distinct phase shifters, were reported more recently \cite{Polino19, Valeri20}. In these devices, the dynamic components were employed to tune the interferometer operation and to implement different phases in the three arms (see Fig.~\ref{fig:reconfigurable}b-c). Such interferometers represent suitable testbeds to benchmark innovative quantum metrology algorithms, allowing in particular for the  simultaneous estimation of two unknown parameters.

Actually, the potentials of reconfigurable integrated circuits would be fully exploited in a completely reconfigurable interferometer realized according to the Reck \cite{Reck94} or Clements \cite{Clem16} scheme, which constitutes a universal linear photonic processor \cite{Carolan15,Taball19} able to implement arbitrary quantum algorithms.
In the femtosecond-laser-written platform, a 4-mode universal circuit following the Clements layout, including twelve distinct thermo-optic phase shifters, was demonstrated in 2018 by Dyakonov et al. \cite{Dyak18}. Here, the resistors were engraved by femtosecond laser ablation on a Nickel-Chromium layer deposited on the fused silica chip, in which the photonic circuit was previously inscribed using the same femtosecond laser. However, operation of that circuit was shown only with classical light.

We ought to note that the use of the thermal phase shifting technology with femtosecond-laser-written circuits may encounter additional practical difficulties with respect to its use in other waveguide platform. In fact, planar technologies typically fabricate the waveguides inside a thin dielectric layer, placed on top of a much thicker silicon substrate which, due to its high thermal conductivity, stabilizes the temperature and quenches heat diffusion far from the heaters. On the contrary, in laser-written devices waveguides are fabricated in a thick glass substrate where heat can slowly diffuse also far from the heater. This decreases the efficiency of the thermal phase shifters, increases cross-talks between neighbouring devices and slows down their time response \cite{Penta21}. However, recent works have shown that thermal efficiency can be enhanced by a careful design of the metallic layer and with an optimized geometry of the microheater \cite{Cecca19}. In addition, the surface of the optical chip can be machined three-dimensionally by the same femtosecond laser to guide the heat diffusion \cite{Chabo17, Cecca20}. In particular, by excavating deep insulation trenches between waveguides, or by embedding the waveguide segment in a suspended glass microbeam, thermal efficiency can be highly increased, and cross-talks diminished, reaching and even surpassing the state-of-art of other waveguide platforms \cite{Cecca20}. Taking advantage of these technological improvements, a new generation of reconfigurable laser-written waveguide circuits will enable more complex quantum photonic experiments in the close future.

\section{Integrated quantum memories}

\begin{figure*}
\centering
\includegraphics[width=\linewidth]{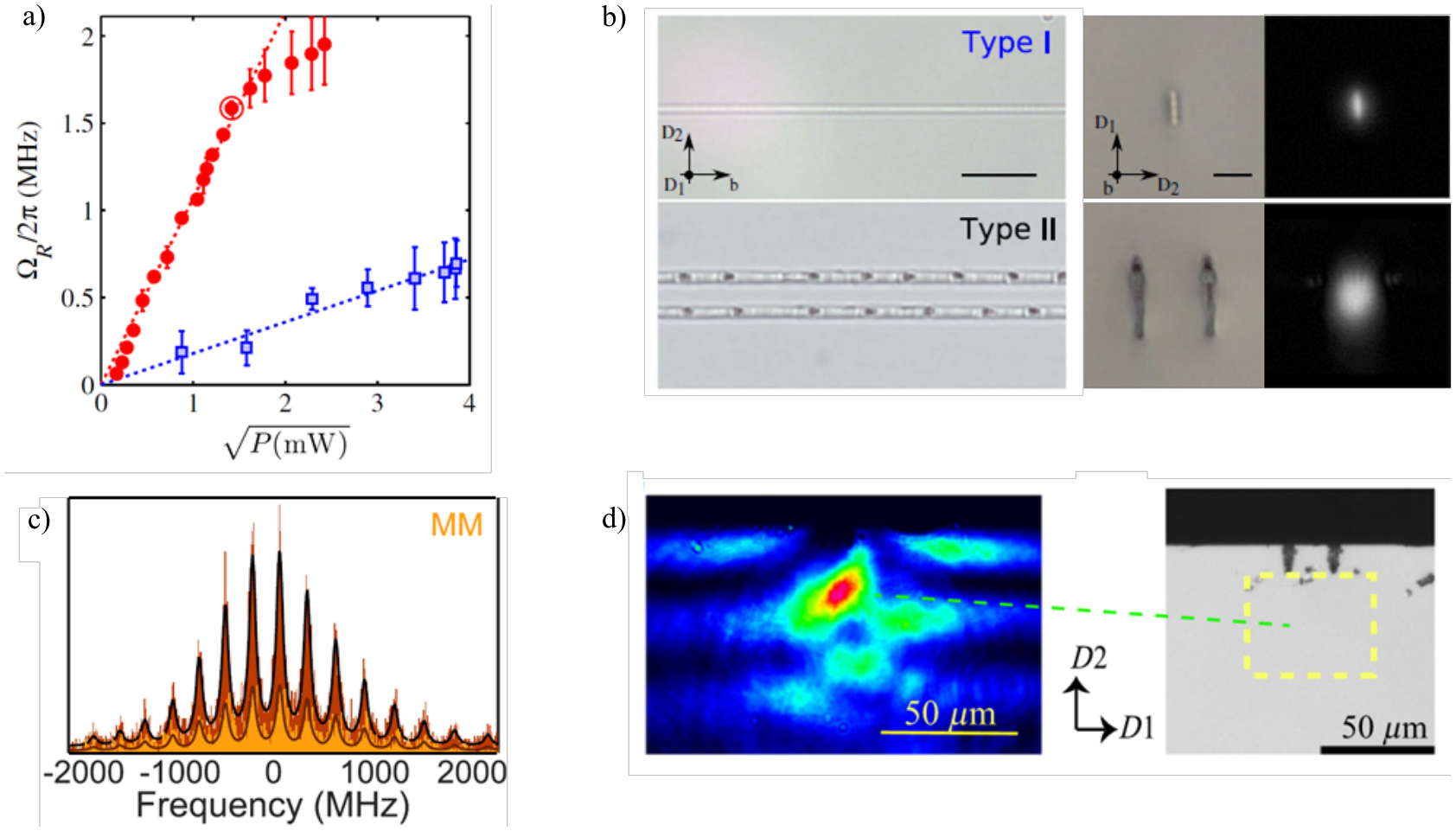}
\caption{\label{fig:memory} (a) Measurement of the Rabi frequency vs pump power performed in bulk Pr$^{3+}$:Y$_2$SiO$_5$ (blue squares) and in a Type II waveguide (red circles) fabricated by FLM in the same material. Experimental details are reported in Ref. \cite{Corr16}. (b) Microscope pictures of the longitudinal profile (scale bar is 20 $\mu$m) and the transverse cross section (scale bar is 10 $\mu$m) of a type-I and a type-II waveguides in Pr$^{3+}$:Y$_2$SiO$_5$, together with the near field images of the corresponding guided modes at 606 nm. See Ref. \cite{Seri18} for a complete comparison between the two platforms. (c) Spectrum of the frequency-multiplexed single photons measured before (brown trace) and after (yellow trace) the storage in a type-I waveguide, as described in Ref. \cite{Seri19}. (d) Transverse cross section of a type-IV waveguide fabricated by FLM in Eu$^{3+}$:Y$_2$SiO$_5$ reported in Ref. \cite{Zhu20}, and the corresponding guided mode profile measured at 580 nm. The panels of this figure contain pictures adapted from Refs. \cite{Corr16}, \cite{Seri18}, \cite{Seri19} and \cite{Zhu20}.}
\end{figure*}

Quantum light storage devices are crucial in the development of large scale quantum technologies. Optical quantum memories (QMs) are, in fact, the building block of most schemes for long distance quantum communications, and play a major role also in photonic quantum computing \cite{Buss13,Hesh16}. Current QM realizations rely on engineered light-matter interfaces, where single photons are absorbed in an optically active medium, and then are re-emitted after a controllable amount of time, while preserving their original quantum state.

Cryogenically-cooled rare-earth-doped crystals are among the most studied systems for developing efficient and long-lived QMs, thanks to the high coherence offered by the dopant ions, the large inhomogeneous broadening of the optical transitions of interest, which allows tailoring the absorption profile by optical pumping, and the possibility to integrate optical waveguides. The latter feature is particularly important in the light of implementing practical QMs in real world applications, which require the simultaneous storage of a very high number of optical modes in a compact and stable fashion. Moreover, the field confinement arising from guided-wave propagation improves the interaction between light and matter, which is highly beneficial for the efficiency of the QM protocols. However, the fabrication of optical waveguides in rare-earth-doped crystals without perturbing their coherence properties is challenging. In this sense, FLM has shown in recent years to be an excellent tool for carrying out this task.

The first pioneering experiment demonstrating a coherent optical memory based on a laser-written optical waveguide was reported in 2016 by Corrielli et al. \cite{Corr16}. A type-II waveguide was inscribed within the bulk of a Pr$^{3+}$:Y$_2$SiO$_5$ crystal, and it was used for the storage and the on-demand recall of classical light pulses at 606~nm employing the atomic frequency comb (AFC) protocol (see Ref. \cite{Afze09} for details on this protocol). In this work it was shown that type-II waveguides fully preserve the coherence properties of the bulk material, since the light propagates in a region which is not directly modified by the laser exposure during the fabrication process. Moreover, the authors explicitly showed that the light matter-interaction in waveguide, quantified as the Rabi frequency of the transition of interest, resulted drastically increased with respect to what measured in bulk under similar focussing conditions (see. Fig.~\ref{fig:memory}a).

In a successive experiment, the same authors have demonstrated that, by tailoring very carefully the laser irradiation parameters, it is possible to inscribe also type-I waveguides in Pr$^{3+}$:Y$_2$SiO$_5$ \cite{Seri18}. In this case, the optical performance in terms of guided-mode dimensions, waveguide propagation losses and bending losses, resulted greatly improved with respect to the type-II counterpart (see. Fig.~\ref{fig:memory}b). In addition, also in this case it was shown that the spectroscopic and the coherence properties of the material were largely unaffected by the inscription process. This enabled to show, for the first time, the quantum storage of heralded single photons in a laser-written waveguide, which outperformed all previous integrated QM demonstrations in terms of storage time, AFC retrieval efficiency and coupling efficiency with the external quantum light source. The same type-I waveguide was then employed for the quantum storage of heralded single photons prepared in the superposition of 15 discrete frequency modes \cite{Seri19}. The strong light-matter interaction in waveguide, in fact, enabled to prepare, by optical pumping, 15 independent AFC frequency bins with a moderate amount of pump power, and to demonstrate the storage and the retrieval of all frequency modes simultaneously. In Fig.~\ref{fig:memory}c) the spectra of the photons before and after the storage are reported. By taking advantage also of the intrinsic temporal multimodality of the AFC protocol, the authors have shown the quantum storage of more than 130 independent optical modes in a single waveguide, thus confirming that FLM offers excellent perspectives towards the practical utilization of QMs in real world scenarios.

More recently, Zhu et al. \cite{Zhu20} have shown that type IV waveguides can be fabricated by FLM in Er$^{3+}$:Y$_2$SiO$_5$. A microscope picture of the type IV waveguide and its guided mode profile are reported in Fig.~\ref{fig:memory}d). Also in this case, it was shown that the coherence properties of the dopant ions inside the waveguide remained unchanged with respect to the bulk crystal. This device was used to demonstrate the coherent storage of classical light pulses at the wavelength of 580 nm adopting the AFC protocol. Since type IV waveguides are fabricated at the crystal surface, this result opens interesting perspectives in coupling laser-written QMs with other surface structures, e.g. coplanar waveguides and/or electrodes, for the coherent driving of the memory operations with external fields. Finally, the same group has demonstrated the inscription of type II waveguides in Er$^{3+}$:Y$_2$SiO$_5$ \cite{Liu20}, and has used it to implement the storage of classical light pulses employing both the AFC protocol and the Revival Of Silenced Echo protocol (ROSE, see Ref. \cite{damo11} for details).

\section{Integrated detection}

The integration of single photon detectors with laser-written photonic circuits is, so far, the less developed capability among those required for realizing a complete photonic quantum architecture. Encouraging results have already been demonstrated for several integrated lithographic platforms, e.g. silicon \cite{pernice12, akhlaghi15, shibata19, martini21}, silicon nitride \cite{ferrari15, kahl15, schuck16} and gallium arsenide \cite{spreng11, mcdonald19, mattioli20} circuits, where superconducting nanowire detectors have been lithographically machined directly on top of the waveguides. Pursuing a similar approach with FLM circuits is more challenging, since laser-written waveguides are buried within the substrate, thus limiting the possibility of interaction between the photons and the detection medium. Further difficulties arise from the non-straightforward applicability of lithographic processes on the glass substrates already machined by FLM. A first attempt in this direction has been demonstrated very recently, in 2021, by Hou and coworkers \cite{hou21}. In this experiment, the authors fabricated, in a borosilicate glass substrate, a set of buried waveguides that bend vertically in a 3D fashion for reaching and crossing the chip top surface. After this step, they performed a sample polishing in order to obtain a sub-nm flatness of the top surface, making it compatible with the process of deposition of a superconducting niobium nitride (NbN) film. Lastly, superconducting NbN nanowires, together with the proper electrical connection pads, were lithographically defined on top of the glass portion where the waveguides cross the surface. This device was then fiber-pigtailed and used as single photon detector, demonstrating an overall detection efficiency of 1.7\%.

In other two recent experiments \cite{heilmann16, Porto18}, FLM optical circuits have been used for the implementation of advanced detection schemes of non-classical states of light. In \cite{heilmann16}, a fused silica FLM chip has been used for sampling the statistics of multiphoton states of light. This device contained a $1 \times 8$ beam divider, fiber-coupled to an array of eight standard avalanche photodiode single photon detectors. In \cite{Porto18}, instead, the authors performed an integrated homodyne tomography of coherent and squeezed light states by means of a laser-written circuit containing a balanced beam splitter and a thermo-optic phase shifter. Although in these experiments the actual photon detecting elements were not integrated within the photonic chip, they show the potential of FLM in developing advanced quantum light analysis. However, incorporating on-chip all the elements for performing a high-efficient and multiplexed single photon detection remains a crucial challenge to overcome for scaling the complexity of laser-written photonic quantum processors, especially for quantum computing and quantum simulation scenarios, where the large number of optical modes makes bulk detection a prohibitive perspective. Therefore, we expect that in the near future this aspect will be studied more and more.

\section{Conclusion and outlook}

FLM is a prominent microfabrication technology for integrated quantum photonics as demonstrated by the large amount of experiments and breakthroughs made possible by this technology. Its role has been important in all domains of quantum information, from computation to simulation, sensing and communication. Its success has been motivated, so far, by its versatility and rapid prototyping capabilities that make it an excellent tool for scientific experiments and proof of principle demonstrations. 

With the increasing effort in developing quantum technologies beyond lab demonstrations, FLM will have to face a new challenge in competing with traditional photolithographic technologies for the development of commercial quantum devices, where scaling of components, packaging and reliability become important features. Fortunately, FLM is following a parallel evolution in becoming an industrially viable technology, with an increasing number of companies that rely on FLM to produce their products. Most likely, FLM has not yet achieved its full maturity, in fact many improvements are still possible, from the development of new physical processes to engineering, and industrialization of the technology.

We believe that the most interesting perspectives for quantum information applications lie in the development of a complete and customizable quantum platform combining many interconnected modules produced by FLM, each dedicated to specific functionalities. The unique three-dimensional capabilities of this technology will be essential to enhance circuit compactness and to explore new paradigms, such as topological photonics \cite{Kremer21}. Finally, the possibility to exploit FLM to combine photonic circuits with microchannels, widely exploited for lab-on-a-chip applications \cite{osellame11}, makes this fabrication technology extremely interesting for the development of integrated quantum sensors and integrated atom and ion traps with embedded photonic circuits.
	
\acknowledgments
 R.O. acknowledges funding from the European Research Council (ERC) under the European Union’s Horizon 2020 research and innovation programme (project CAPABLE — Grant agreement No. 742745 - www.capable-erc.eu).

A.C. acknowledges funding from the Ministero dell'Istruzione, dell'Universita e della Ricerca (http://dx.doi.org/10.13039/501100003407, PRIN 2017 programme, QUSHIP project - id. 2017SRNBRK).

\end{document}